\documentclass[twocolumn]{article}
\usepackage{graphicx} 
\usepackage{amsmath}
\usepackage[braket, qm]{qcircuit}
\usepackage{caption, subcaption}
\usepackage{hyperref, cite, listings, xcolor, float}
\title{ONDA: A High-Level Quantum Programming Language with Sequential Execution and Conditional Branching}
\author{Francesco Junior De Gregorio}
\date{September 2024}
\definecolor{keyword_color}{RGB}{184, 11, 40} 
\definecolor{id_color}{RGB}{7, 56, 163}
\definecolor{comment_color}{RGB}{98, 105, 120}
\definecolor{not_color}{RGB}{112, 90, 28}
\lstdefinelanguage{Assembly}{
    keywords={addi, add, caddi, caiu},  
    sensitive=true,                            
    comment=[l]{\#},
    morekeywords=[2]{not},
    keywordstyle=\color{keyword_color},                 
    keywordstyle=[2]\color{not_color},
    numberstyle=\color{number_color},                   
    identifierstyle=\color{id_color},
    commentstyle=\color{comment_color}
}

\lstdefinelanguage{PseudoCode}{
    keywords={if, else, addi, swap, flip, increment},  
    keywordstyle=\bfseries,  
    comment=[l]{\#},         
    commentstyle=\color{gray!60}\itshape,  
    moredelim=[is][\itshape]{\{\{}{\}\}}, 
    moredelim=[s][\itshape]{...}{...},    
    basicstyle=\ttfamily\footnotesize,  
    showstringspaces=false,
    breaklines=true,
    tabsize=4
}

\lstdefinestyle{C-style}{
    backgroundcolor=\color{gray!10}, 
    keywordstyle=\color{blue},
    commentstyle=\color{green!50!black},
    stringstyle=\color{red},
    basicstyle=\ttfamily\footnotesize,
    breaklines=true,
    tabsize=4
}

\begin{document}

\maketitle
\begin{abstract}
    This paper introduces ONDA, a quantum programming language designed to significantly simplify quantum programming by providing multiple abstraction layers similar to those found in classical computing. Unlike traditional quantum programming languages, which primarily focus on circuit construction, ONDA compiles into quantum instructions executed autonomously by specialized quantum hardware, eliminating the need for classical assistance. The proposed architecture uniquely supports sequential execution, branching, and the direct implementation of classical control structures such as conditional statements and loops (e.g., if, do-while) entirely within the quantum domain. By leveraging a quantum microarchitecture that autonomously processes compiled instructions, ONDA facilitates the intuitive implementation of high-level quantum algorithms, reducing complexity and broadening access to quantum programming to promote wider adoption and accelerate advancements in quantum computing.

\end{abstract}
\tableofcontents
\section{Introduction} \label{sec:introduction}
Quantum computers in the current NISQ era are noisy, error-prone, and incapable of executing many operations before decoherence occurs. However, the field is experiencing tremendous growth and promising road maps guiding future advancements. In recent years, remarkable achievements have been made \cite{ibm_evidence_of_utility, Machine_learning, QAOA, Quantum_supremacy}, and small-scale quantum computers are increasingly becoming available for public research. This growing accessibility highlights the need for expertise in the field. In response, companies have begun to reduce the knowledge barrier for programming quantum computers and have introduced educational programs. However, these efforts are still aimed at specialized individuals with a scientific background.

Unlike classical programming, which has become widely accessible, quantum programming remains limited due to its low-level interface requiring users to build circuits directly. One key factor behind the exponential growth of classical computing is the ability to create multiple layers of abstraction, allowing programmers to work at higher levels without needing to understand every underlying detail. This work proposes an architecture that introduces similar layers of abstraction for quantum computing, enabling sequential execution and branching through a high-level quantum programming language.

Quantum programming languages have functioned as interfaces for constructing quantum circuits, classically relying on external computation. In contrast, this paper presents ONDA, a programming language, and an architecture that allow programs to be written in a structured textual format, classically compiled, and executed autonomously on a quantum architecture without reliance on a classical computer. Notably, while other programming languages have depended on classical computers for conditional operations, ONDA eliminates the need for classical conditional processing. This paper explores both the hardware architecture and the corresponding software interface designed for its programming, it is organized as follows:
Section \ref{sec:Execution_Model_and_Architectural_Overview} introduces the execution model and architectural design of the proposed quantum microarchitecture. Section \ref{sec:software} defines the Quantum Instruction Set Architecture (ISA) and assembly language, outlining how classical programming constructs are implemented within a unitary quantum framework. Section \ref{sec:ONDA_programming_language} presents ONDA, a high-level quantum programming language designed to simplify quantum programming, providing an overview of its syntax and quantum-specific operations.

Finally, Section \ref{sec:future_outlooks} discusses potential future advancements, while Section \ref{sec:Conclusion} summarizes key contributions and outlines the broader implications of this work.
\begin{figure}
    \centering
    \includegraphics[width=\linewidth]{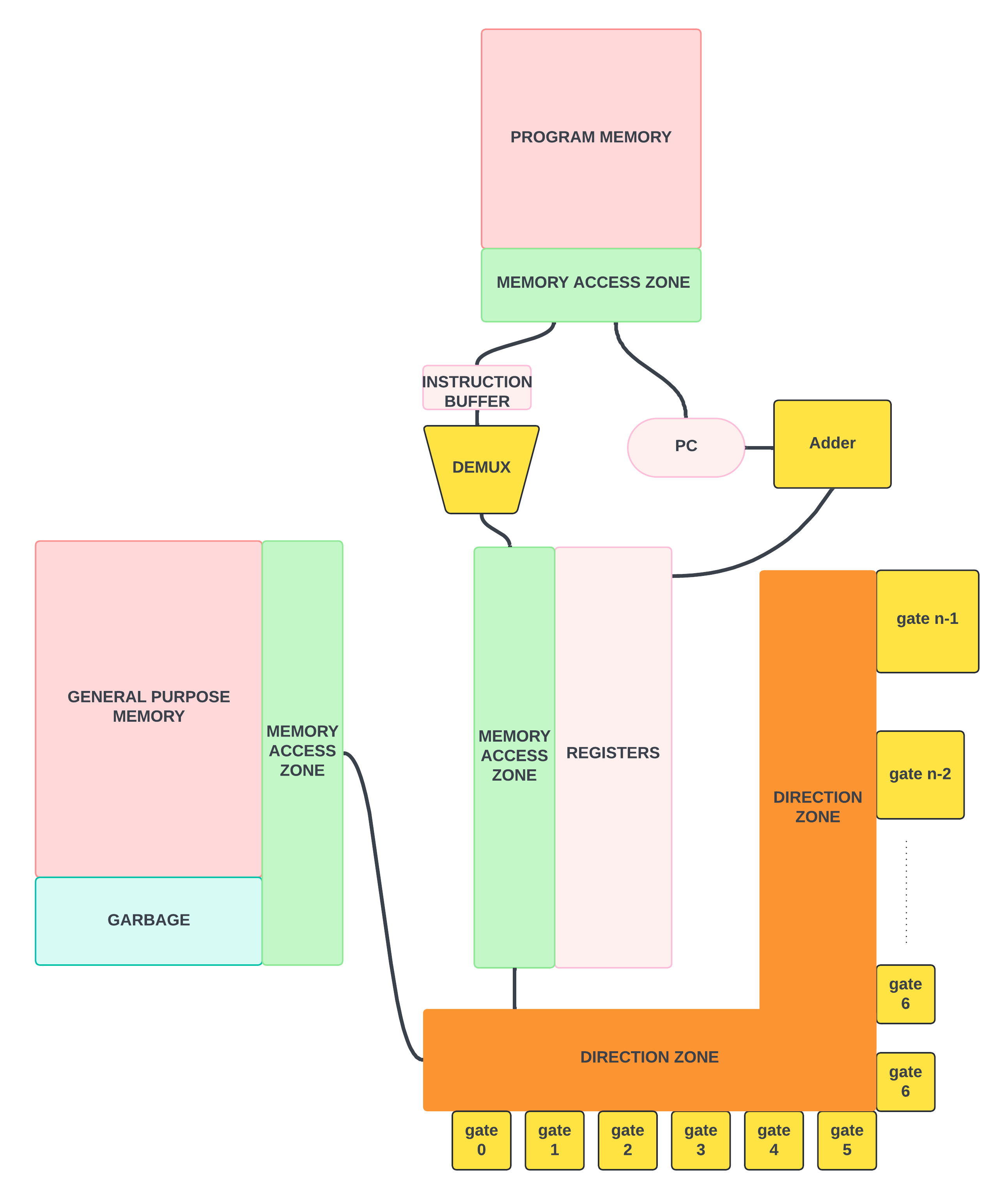}
    \caption{A schematic representation of the quntum architecture.}
    \label{fig:Architecture_map}
\end{figure}

\section{Execution Model and Architectural Overview}\label{sec:Execution_Model_and_Architectural_Overview}
\subsection{General Execution Model}
\paragraph{}
The architecture described here is constructed on several perfect qubits, numbering on the order of magnitude greater than those currently available. One should envision a plane densely populated with qubits, each connected to two or three other qubits, thereby forming a grid that will henceforth be referred to simply as a chip. Classical chips perform computations by repeatedly running a circuit for many cycles; similarly, in this architecture, a unitary operation $S$ is executed repeatedly. This operation is analogous to the “clock cycle” of classical chips.

\subsection{Classical Computational Architecture}
Prior to discussing the quantum computational architecture, it is reviewed a simplified model of the classical architecture. It should be acknowledged that this description represents an idealized scenario; contemporary chips incorporate numerous additional features and complexities that extend beyond the scope of this discussion.
\paragraph{}
Classically, a program—a list of instructions—is loaded into program memory, and a pointer tracks the index of the current instruction being executed. At each cycle, the instruction pointed to, composed simply of 64 bits, is copied into the control unit (CU). The CU then decodes the 64-bit instruction, extracting \textit{opcode} and \textit{operands}\footnotemark{}, and signals the arithmetic logic unit (ALU) to execute the instruction. Once the ALU completes its computation, the result is stored in the registers specified by the operands. In cases of memory operations, such as loading a value, the operation is redirected to memory in a similar way. To conclude the cycle, the program counter (PC) is updated. It is noteworthy that, to implement more complex control logic such as if and for statements, the PC is dynamically modified to point to the appropriate instruction.
To emulate classical operations on a quantum computer, each step must be unitary and reversible, thus requiring careful adaptation of traditional classical operations.

\footnotetext{An instruction comprises opcode and operands. The former specifies the operation to be executed, while the latter designates the registers on which the operation is performed.}
\subsection{Parallel Quantum Computation}
This paper will focus exclusively on quantum scenarios in which the machine is in a basis state. Nevertheless, the computational scheme remains valid for states in superposition. Specifically, different computations can be performed in parallel without unintended interference between states. Let $\ket{a}, \ket{b} \in \Sigma$ where $\Sigma$ represent the set of all basis states. Up to normalization:
\begin{gather*}   
    S^i\ket{a} = \ket{\psi_1} \\
    S^i\ket{b} = \ket{\psi_2}\\
    \ket{\phi} = \ket{a} + \ket{b}
    \\S^i\ket{\phi} = \ket{\psi_1} + \ket{\psi_2}
\end{gather*}

\subsection{Quantum Processing Unit Architecture} \label{sec:quantum_architecture}
The quantum chip, illustrated in Fig \ref{fig:Architecture_map}, includes regions analogous to its classical counterpart. This subsection details a sequence of reversible operations comprising one cycle on the chip. Clearly, the operation $S$ is defined as the product of these operations.
\paragraph{}
Initially, at the start of computation, a program—represented as a sequence of qubits initialized to $\ket{0}$ and $\ket{1}$—is loaded into the program memory, and the Program Counter is initialized to 0. Employing a QRAM-like approach \cite{classic_qram}, the instruction indicated by the PC is fetched into an instruction buffer (implemented as a simple 64-qubit register). Subsequently, the opcode is decoded with a quantum demultiplexer, producing $n$ \textit{activation qubits}, each indicating whether a specific instruction should be executed, where $n$ is the total number of instructions in the architecture. In a basis-state, exactly one activation qubit is set to 1 per cycle. Concurrently, the values of the registers specified by the operands are transferred into temporary buffers, with access to these registers performed again in a QRAM-like manner. Following this, the values of the operands and the activation qubits are transmitted to the Arithmetic Logic Unit; the specifics of this transmission process are discussed in Section \ref{sec:trasmission_mechanism}. Within the ALU, each instruction is assigned a distinct region with dedicated input and output buffers. Operand values and activation qubits traverse the ALU, and at the entry of each instruction, the operand values are conditionally swapped with the ancilla qubits in the input/output buffers\footnotemark, with the activation qubit of the instruction serving as the control qubit for the controlled swap. In a basis state, the input and output buffers of all instructions contain ancilla qubits, except for those of the currently executing instruction, whose ancillas remain in the bus that traverses the ALU.
\footnotetext{In this approach, outputs are written to qubits originally sourced from the registers.}

Subsequently, all instructions within the ALU are executed (e.g., arithmetic operations like addition and subtraction). Afterwards, the ALU performs a sequence of reversal operations that restore the ancilla qubits to their initial state while ensuring the input and output qubits remain unchanged. To accomplish this, one may utilize the general circuit depicted in Fig \ref{fig:instruction_implementation}, or implement more optimized algorithms such as \cite{Carry_Save_Arithmetic, qft_based_addition}, that avoid the necessity of output qubits initialized to zero. Thereafter, by performing the inverse of all previous operations (excluding those within the ALU) up to and including reinserting the instruction into program memory, the Quantum CPU is restored nearly to its initial state, with differences only in the processed qubits. Finally, by emulating the classical architecture, the PC would be incremented by one.

However, this approach inherently constrains the complexity of executable logic. Within this scheme, constructs such as \textit{if} and \textit{do-while} statements are not feasible since computation cannot modify the PC during a cycle; altering the PC while an instruction resides in the instruction buffer would impede reinserting the instruction into program memory, as its original address is required. To overcome this limitation, this study proposes the implementation of an "update register." This register operates similarly to standard registers, enabling computational operations to modify its contents; however, at the conclusion of each cycle (after all reversal operations are executed), its value is added to the PC instead of incrementing the PC by one. This approach permits the realization of control flow constructs, including \textit{if-else} and \textit{do-while} statements, which are discussed in further detail in Section \ref{sec:control_flow_implementation}.

\begin{figure}
    \centering
    \[
\Qcircuit @C=0.8em @R=0.8em @!R {
    & \lstick{input: \ket{x}_{1}} & \multigate{9}{\mathrm{U}} & \qw & \qw & \qw & \multigate{9}{\mathrm{U^{\dag}}} & \qw & \rstick{\ket{x}_{1}} \qw \\
    & \vdots & & \vdots & & \vdots & & & \vdots \\
    & \lstick{input: \ket{x}_{n}} & \ghost{\mathrm{U}} & \qw & \qw & \qw & \ghost{\mathrm{U^{\dag}}} & \qw & \rstick{\ket{x}_{n}} \qw \\
    & \lstick{ancilla: \ket{0}_{1}} & \ghost{\mathrm{U}} & \qw & \qw & \qw & \ghost{\mathrm{U^{\dag}}} & \qw & \rstick{\ket{0}_{1}} \qw \\
    & \vdots & & \vdots & & \vdots & & & \vdots \\
    & \lstick{ancilla: \ket{0}_{a}} & \ghost{\mathrm{U}} & \qw & \qw & \qw & \ghost{\mathrm{U^{\dag}}} & \qw & \rstick{\ket{0}_{a}} \qw \\
    & \lstick{ancilla: \ket{0}_{1}} & \ghost{\mathrm{U}} & \ctrl{4} & \qw & \qw & \ghost{\mathrm{U^{\dag}}} & \qw & \rstick{\ket{0}_{1}} \qw \\
    & \lstick{ancilla: \ket{0}_{2}} & \ghost{\mathrm{U}} & \qw & \ctrl{4} & \qw & \ghost{\mathrm{U^{\dag}}} & \qw & \rstick{\ket{0}_{2}} \qw \\
    & \vdots & & & & & & & \vdots \\
    & \lstick{ancilla: \ket{0}_{m}} & \ghost{\mathrm{U}} & \qw & \qw & \ctrl{4} & \ghost{\mathrm{U^{\dag}}} & \qw & \rstick{\ket{0}_{m}} \qw \\
    & \lstick{output: \ket{0}_{1}} & \qw & \targ & \qw & \qw & \qw & \qw & \rstick{\ket{f(x)_{1}}} \qw \\
    & \lstick{output: \ket{0}_{2}} & \qw & \qw & \targ & \qw & \qw & \qw & \rstick{\ket{f(x)_{2}}} \qw \\
    & \vdots & & \vdots & & & & & \vdots \\
    & \lstick{output: \ket{0}_{m}} & \qw & \qw & \qw & \targ & \qw & \qw & \rstick{\ket{f(x)_{m}}} \qw \\
}
    \]
    \caption{Quantum circuit for implementing any classical operation. The circuit employs the quantum counterparts of classical gates within an operation $U$ utilizing $a + m$ ancilla qubits, where $a$ store temporary values and $m$ hold the output. The output is then copied onto $m$ zero-initialized qubits via CNOT gates, followed by the reversal of $U$ to restore the input and ancilla qubits while preserving the output on $m$ qubits.}
    \label{fig:instruction_implementation}
\end{figure}
\subsection{Quantum Data Transmission Mechanisms}\label{sec:trasmission_mechanism}
The QPU comprises many error-corrected qubits that remain stationary while the information is transmitted. Various methods can be employed to transmit information between different areas, two of them are discussed here.

\subsubsection{CNOT-Based Copy Mechanism}

An efficient method for moving information is to apply a series of CNOT gates to ancilla qubits initialized to 0 starting from the qubits that have the information, as in Fig \ref{fig:Cnot-copyl}. During computation, the bus cannot be reused, as all the qubits are occupied by the values of the original data. Once the information reaches its destination, it can only be read; any modification would preclude the ability to reverse the operation. This method of information transmission can be referred to as "copy," as one cannot modify the original value but rather possesses a view of it.

\subsubsection{Swap-Based Reference Mechanism}

Using a structure similar to the previous method, a series of swap gates can also be used. Although this approach is slower, three times slower than the CNOT copy, it provides a greater degree of freedom. The bus can be reused while the data is being processed by the computation. Furthermore, one can modify the data or swap it with other information to execute move operations. This method of information transmission can be referred to as "reference," as it allows for manipulation of the actual values stored in the source qubits.

\begin{figure}
    \centering
    \[
    \Qcircuit @C=1.0em @R=0.8em @!R { \\
	 	\nghost{{\;\;\;\;\;\;data\;:\;\ket{x}} :  } & \lstick{{\;\;\;\;\;\;data\;:\;\ket{x}}  } & \ctrl{1} & \qw & \qw & \qw & \qw & \qw\\
	 	\nghost{{\;\;ancillla_0\;:\;\ket{0}}_{0} :  } & \lstick{{\;\;ancillla_0\;:\;\ket{0}}  } & \targ & \ctrl{1} & \qw & \qw & \qw & \qw\\
	 	\nghost{{\;\;ancillla_0\;:\;\ket{0}}_{1} :  } & \lstick{{\;\;ancillla_1\;:\;\ket{0}}  } & \qw & \targ & \ctrl{1} & \qw & \qw & \qw\\
	 	\nghost{{\;\;ancillla_0\;:\;\ket{0}}_{2} :  } & \lstick{{\;\;ancillla_2\;:\;\ket{0}}  } & \qw & \qw & \targ & \ctrl{1} & \qw & \qw\\
	 	\nghost{{\;\;ancillla_0\;:\;\ket{0}}_{3} :  } & \lstick{{\;\;ancillla_3\;:\;\ket{0}}  } & \qw & \qw & \qw & \targ & \qw & \qw\\
\\ }
    \]
    \caption{Cnot copy circuit representation}
    \label{fig:Cnot-copyl}
\end{figure}

\section{Quantum Instruction Set Architecture and Assembly Language}\label{sec:software}
ONDA is a programming language with syntax similar to that of the C programming language. It compiles into an intermediate quantum assembly language comprising a small set of low-level instructions. These instructions can be directly implemented using quantum gates and are reversible as defined in Section \ref{sec:quantum_architecture} 

The limited number of instructions in the intermediate assembly language is intentional, providing flexibility to adapt them according to specific application requirements or to simplify their practical implementation. In contrast, ONDA itself aims to serve as a universal language, enabling the expression of quantum algorithms independently of the specifics of the underlying instruction set architecture.

This section addresses methods for implementing classical programming constructs, such as conditional jumps, within the constraints of unitary quantum operations.

\subsection{Syntax and Program Structure}
Programs stored in program memory are encoded in binary format. Assembly languages are representations of these programs in a human-readable format, in which nearly each keyword corresponds directly to an operation executed by the chip. Assembly code is classically translated (or assembled) into a sequence of bytes, subsequently loaded into program memory. An assembly program consists of multiple lines, each representing an individual instruction. The general format of an instruction is:
\[
 \langle \text{instruction} \rangle \langle \text{inputs} \rangle
\]
$\langle \text{instruction} \rangle$ acts as the operation identifier. $\langle \text{inputs} \rangle$  comprises a series of tokens representing either registers (identified by their aliases as outlined in \autoref{tab:registers_alises}) or immediate numerical values. A representative example of assembly code is provided below:
\begin{lstlisting}[language=assembly]
    addi $t1, 6   
    add  $t2, $t1 
\end{lstlisting}
In this program the first instruction add 6 to \$t1\footnotemark, Then the second instruction add \$t1 to \$t2.
The complete set of assembly instructions into which ONDA is compiled consists of 26 instructions and is detailed comprehensively in \cite{demialenzo_quantum_isa}.

\footnotetext{In this paper, any reference to a register alias actually refers to the value stored in the register.}
\begin{table*}[ht]
    \centering
    \caption{Registers aliases}
    \begin{tabular}{|c|c|c|} 
        \hline 
        \textbf{REGISTERS INDEXES} & \textbf{REGISTERS ALIASES} & \textbf{REGISTERS NAMES} \\ 
        \hline 
        0                 & \$zero           & 0 register                 \\ 
        \hline
        1                 & \$ur             & update register             \\ 
        \hline
        2                 & \$tur            & temporary update register   \\ 
        \hline
        3-4               & \$v0-\$v1          & return values               \\ 
        \hline
        5-8               & \$a0-\$a3          & argument regs   \\ 
        \hline
        9-18              & \$s0-\$s9        & saved regs                   \\ 
        \hline
        19-27             & \$t0-\$t8          & temporary regs  \\ 
        \hline
        28                & \$grp            & garbage pointer             \\ 
        \hline
        29                & \$sp             & stack pointer               \\ 
        \hline
        30                & \$fp             & frame pointer               \\ 
        \hline
        31                & \$ra             & return address              \\ 
        \hline 
    \end{tabular}
    \label{tab:registers_alises} 
\end{table*} 
\subsection{Control Flow Implementation in Assembly}\label{sec:control_flow_implementation}
\subsubsection{If statement}
The two branches of an if statement are implemented using conditional additions to the update register (UR). When the condition for a jump evaluates to true, the jump occurs through two sequential operations: firstly, the length of the instruction block to be skipped is added to the UR, directly causing the jump; secondly, the same value is subtracted from the UR to reset it to its original state after the program counter has bypassed the intended instruction block.
\begin{lstlisting}[language=Assembly]
    # eval condition
    caddi not cond $ur +x
    # ...true body with length x...
    caddi not cond $ur -x
    caddi cond $ur +y
    # ...false body with length y..
    caddi cond $ur -y
    # deval condition 
\end{lstlisting}
The operation caddi is the conditional add immediate instruction, it adds the immediate value to the second register if the first quibit of the first register is 1. 
\subsubsection{Do while loop}
The algorithm implementing a do-while loop within the ONDA assembly language utilizes two temporary update registers specifically to manage conditional jumps that return execution to the loop's initial instruction. These registers cyclically assume the values $-x+1$, $1$, and $0$, enabling conditional jumps to either return to the start of the loop, advance through the loop body, or control instruction repetitions. 
\begin{lstlisting}[language=PseudoCode]
    x = BODY_LENGTH + 1
    tur1 = 1
    tur2 = -x+1
    counter = 0
    activation = False
    # Two cases swap (tcs)
    # (it is one single asm instruction)
    {
    if counter % 2 == activation:
        swap ur tur1
    else:
        swap ur tur2
    flip activation
    }
    # Two cases add immediate (tcai)
    if counter % 2 == 0:
        addi tur2 x
    else
        addi tur1 x
    
    # set tur1 = 0 only 
    # at the first iteration
    if counter == 0
        addi tur1 -1
    
    .....
    body
    .....
    
    increment counter
    if arbitrary condition:
        addi ur -x
\end{lstlisting}

The correctness of this approach can be verified through an analysis of the cases presented in Table \ref{tab:do_while}, which demonstrates that the given sequence of instructions accurately reflect the semantics of a standard do-while loop.

Although only four iterations are explicitly shown, it is important to note that the loop counter consistently changes every iteration, while the two temporary registers alternate values cyclically with a period of two iterations. Since the algorithm's logic depends solely on the parity of the loop counter, these illustrative iterations are sufficient to establish its correctness for any number of iterations.

\renewcommand{\arraystretch}{1.2} 
\setlength{\tabcolsep}{3pt} 

\begin{table*}
    \centering
    \resizebox{\textwidth}{!}{ 
        \begin{tabular}{|l|ccccc|ccccc|ccccc|ccccc|}
            \hline
            & \multicolumn{5}{c|}{\textbf{ITERATION 1}} 
            & \multicolumn{5}{c|}{\textbf{ITERATION 2}} 
            & \multicolumn{5}{c|}{\textbf{ITERATION 3}} 
            & \multicolumn{5}{c|}{\textbf{ITERATION 4}} \\
            \hline
            & $\$ur$ & tur1 & tur2 & activation & counter 
            & $\$ur$ & tur1 & tur2 & activation & counter 
            & $\$ur$ & tur1 & tur2 & activation & counter 
            & $\$ur$ & tur1 & tur2 & activation & counter \\
            \hline
            initialization 
            & 1 & 1 & -x+1 & FALSE & 0 
            &  &  &  &  & 
            &  &  &  &  &  
            &  &  &  &  &   \\
            
            tcs 
            & 1 & 1 & -x+1 & TRUE & 0 
            & 0 & -x+1 & 1 & FALSE & 1 
            & 0 & 1 & -x+1 & FALSE & 2 
            & 0 & -x+1 & 1 & FALSE & 3 \\
            
            tcs (only if previous $\$u$ was 0) 
            &  &  &  &  &  
            & 1 & -x+1 & 0 & TRUE & 1  
            & 1 & 0 & -x+1 & TRUE & 2  
            & 1 & -x+1 & 0 & TRUE & 3  \\
            
            tcai 
            & 1 & 1 & 1 & TRUE & 0 
            & 1 & 1 & 0 & TRUE & 1 
            & 1 & 0 & 1 & TRUE & 2 
            & 1 & 1 & 0 & TRUE & 3 \\
            
            if counter=0 tur1 -=1 
            & 1 & 0 & 1 & TRUE & 0 
            & 1 & 1 & 0 & TRUE & 1 
            & 1 & 0 & 1 & TRUE & 2 
            & 1 & 1 & 0 & TRUE & 3 \\
            
            ..BODY.. 
            &  &  &  &  &  
            &  &  &  &  &  
            &  &  &  &  &  
            &  &  &  &  &  \\
            
            increment counter 
            & 1 & 0 & 1 & TRUE & 1 
            & 1 & 1 & 0 & TRUE & 2 
            & 1 & 0 & 1 & TRUE & 3 
            & 1 & 1 & 0 & TRUE & 4 \\
            
            caddi loop\_cond \$ur - x 
            & -x+1 & 0 & 1 & TRUE & 1 
            & -x+1 & 1 & 0 & TRUE & 2 
            & -x+1 & 0 & 1 & TRUE & 3 
            & -x+1 & 1 & 0 & TRUE & 4 \\
            \hline
        \end{tabular}
    }
    \caption{Table illustrating iterations and corresponding variable values during the execution of the `do-while` loop. When the UR variable holds a negative value, adding it to the PC causes the execution flow to revert to the first instruction of the loop. Conversely, when the UR variable is set to zero, the PC remains unchanged, resulting in the repeated execution of the same instruction.}
    \label{tab:do_while}
\end{table*}

\subsubsection{Function call}
The following pseudocode illustrates the jump logic associated with a function call. The algorithm use of temporary update register, along with arithmetic operations and register swapping to enables accurate calculation and management of instruction addresses of the function (qfun), the return statemen (qret), and function call (qcall).

\begin{lstlisting}[language=PseudoCode]
    qfun: # line_num = f_add
        swap ur tur 
        .....
        body
        .....
    qret: 
        tur *= -1
        tur -= fun_len
        # tur = c_add - f_add - fun_len 
        swap ur tur
    
    qcall:
        tur = f_add - c_add 
        swap ur tur
        .....
        code
        .....
        

\end{lstlisting}
Where c\_add, f\_add and fun\_len are, respectively, the address of the function call, the address of the function and the length of the function. 

\subsection{Quantum Garbage Management and Memory Handling}\label{sec:quantum_garbage}
Quantum Instruction Set Architectures present unique challenges that classical ISAs do not face. Traditional operations such as "save word," "load word," and "jump" are not unitary transformations and, therefore, cannot be directly implemented within a quantum computing framework. Despite these constraints, many quantum computing procedures require the use of qubits initialized to the $|0\rangle$ state. Furthermore, certain computed values are only required for specific stages of a computation and can be discarded once they are no longer needed. However, resetting a qubit to $|0\rangle$ is inherently non-unitary and cannot be performed without introducing measurement-induced interference or collapsing the quantum state. To mitigate this issue, a  memory management approach is introduced, using a specialized 0-initialized memory region called "garbage". This memory structure maintains a garbage stack, with its top element tracked by a designated garbage pointer (\$grp). When a value needs to be reset to \(|0\rangle\), it is swapped with the top element of the stack, and the garbage is updated accordingly.  

\section{ONDA: High-Level Quantum Programming Language}\label{sec:ONDA_programming_language}
The ONDA programming language is designed to provide an environment in which programmers can disregard garbage handling and focus solely on writing code. ONDA is a programming language similar to C and shares many of its features; therefore, the classical aspects will be briefly discussed while placing greater emphasis on the unique quantum features.
\subsection{Classical Constructs and Syntax Overview}
Operators +=, -=, inline expression like $2 + a * (b/4 - 1)$ are supported without garbage creation, while  *=, /=, \%= , re-assignation ($a = b$) create garbage. Function calls and definitions are equal to C. For further details about the syntax see \cite{demialenzo_quantum_isa}.
\begin{lstlisting}[language=C, style=C-style]
    // ONDA CODE EXAMPLE
    int a; // same as a = 0
    int b = 3;
    float c; // same as c = 0.0
    float d = 4.5;
    int arr1[32]; 
    float arr2[32];
    a += 2; 
    a += b; 
    a = (b + 4) * 2;
    arr[a] = 2;
    if(condition1){
    // do something 
    }else{
    // do something else
    }  
    do{
    	// ...computation...
    }while(condition2);
    
\end{lstlisting}
\subsection{Quantum-Specific Operations and Expressions}
In ONDA, three quantum operations are defined: X, Z, and H.

The X operation is represented by the $\sim$  operator and functions identically to its classical counterpart in the C programming language.
\begin{lstlisting}[language=C, style=C-style, literate={~}{$\sim$}1]
    // a = int max
    unsigned int a = ~0; 

    // a = 0
    a = ~a;
\end{lstlisting}    
The $Z$ operation is denoted by the "$\#$" operator. In an expression such as $a \# b$, a $Z$ gate is applied to the first $b$ qubits of the value $a$, starting from the least significant bit (LSB). Similarly, the $H$ operation is denoted by the "$@$" operator. In an expression such as $a @ b$, a Hadamard gate is applied  to the first b qubits of the value a. By combining these operators, one can construct a wide range of programs.
\begin{lstlisting}[language=C, style=C-style]
    int a = 0 @ 6; 
    a @= 2; 
    a += 1;
    a #= 1;
\end{lstlisting}
At the start, the state of $a$ is an equal superpositon of all the values from 0 to 64. Then it become a superpositon of all the values from 0 to 64 that are multiples of 4. Since the  LSB is 0, the instruction $a += 1$ flips the last qubit. The last instruction perform a $Z$ gate on the LSB kicking a phase into the system.

\subsection{Example Implementations in ONDA}
\subsubsection{Deutsch-jozsa algorithm}
The Deutsch-Jozsa algorithm implementation in ONDA involves declaring and initializing variables, applying the Hadamard transform, invoking the oracle function ("secret function"), applying the Hadamard transform again, and finally measuring the quantum state using a print statement.
\begin{lstlisting}[language=C, style=C-style]
int main()
{
    int n = ....;
    int input_num = 0;
    int aux_num = 1;
    input_num @= n;
    aux @= 1; 
    secret_fun(input_num, aux);
    input_num @= n;
    print input_num
    return;
}
\end{lstlisting}
\subsubsection{Order finding algorithm}
The order-finding algorithm is implemented utilizing a do-while loop. Following the initialization phase, the algorithm iteratively examines each qubit individually. Specifically, the value of each qubit is checked through a bitwise operation ((num \texttt{<}\texttt{<} m) \& 1). If the expression returns a value of one, the auxiliary qubits are multiplied directly by a power of $a$ (the pow of $a$ is computed in logaritmic time). Subsequently, a Quantum Fourier Transform is applied to the resultant state. Finally, a measurement is performed, producing an output that requires classical post-processing.

\begin{lstlisting}[language=C, style=C-style]
int pow(int base, int exp){ ... }
int QFT(int arg, int base){ ... }
int main()
{
    int a = ...;
    int m = ...;
    int num = 0; 
    int aux = 1;
    int i = 0;
    num @= m;
    do{
        if((num << i) & 1){
            aux *= pow(a, 1<<i);
    }(i < m);

    QFT(num, 2<<m);
    print num;
    return;
}   
\end{lstlisting}
\subsubsection{Grover's search}
This example illustrates the implementation of Grover's search algorithm applied to an unstructured database \cite{grover}. Following the initial state preparation, the Grover operator is repeatedly applied using a do-while the exact number of times. To induce the appropriate phase inversion for targeted states, a controlled-Z gate is applied to an ancillary qubit initialized in the state \(|1\rangle\) (represented programmatically by the statement \texttt{one \#= 1}). Then the implementation adheres closely to the original specification of the algorithm.
\begin{lstlisting}[language=C, style=C-style]
int arr = {...};
int main()
{
    float pi = 3.141592;
    int arr_length = 256;
    int index = 0;  
    int searched_numeber = 36;
    int iteration = 0;
    int one = 1;
    
    index @= 8; /* super impose all the 
                indexes from 0 to 255*/
    do{
      if(arr[index] == searched_number){
        // kick back the phase 
        one #= 1; 
      }
      index @= 8;
      if(index == 0){
        // kick back the phase
        one #= 1; 
      }
      index @= 8;
      iteration += 1;
    }while(iteration < (pi / 4) * sqrt(arr_length));
    
    print index;
    return;
}
/* the most probable measure is 
the index of 36*/
\end{lstlisting}
\section{Future outlooks}\label{sec:future_outlooks}
Modern architectures are significantly more complex and optimized compared to the proposed quantum model. However, in the future, many classical features may become translatable into their quantum counterparts. Moreover, entirely novel approaches could emerge by leveraging the distinctive properties of quantum mechanics. For instance, new bus architectures based on quantum teleportation protocols \cite{high_dimensional_teleportation} could be developed.

Further advancements are also expected in the hardware-software co-design of quantum processing units. One promising direction is the inclusion of unused opcodes, enabling programmers to define and dynamically invoke custom quantum instructions within the architecture.

Additionally, this proposal presents numerous opportunities for both hardware and software optimizations, as well as standardization efforts across various domains. For example, determining the optimal number $i$ of iterations for the step operation $S$ before measurement $M$ remains an open question. Future research can further explore high-level quantum programming languages, such as ONDA, to identify new application domains and enhance the broader software ecosystem of quantum computing.
\section{Conclusion}\label{sec:Conclusion}
This paper introduces ONDA, a quantum programming language and microarchitecture designed to enable high-level quantum programming with sequential execution and branching. Unlike conventional quantum programming models that rely on classical computation for control flow, ONDA operates autonomously within a quantum framework, eliminating the need for external classical processing.

The proposed quantum microarchitecture defines a structured execution model where instructions stored in program memory are sequentially fetched, decoded, and executed within a Quantum Processing Unit. The architecture includes mechanisms for data transmission, control flow management, and arithmetic operations, all implemented using reversible quantum logic. Additionally, a convention for constructing gates and executing instructions in a quantum ALU has been established.

A Quantum Instruction Set Architecture is defined, detailing a set of low-level instructions that support fundamental computational tasks while preserving unitarity. Algorithms for implementing classical programming constructs such as conditional branching, loops, and function calls within this framework are explored. To address the challenges of quantum memory management, a garbage collection strategy using a specialized register is proposed.

The ONDA programming language, designed as a high-level abstraction over the ISA, provides a structured syntax similar to classical programming languages while introducing quantum-specific operations. Several example implementations, including the Deutsch-Jozsa algorithm, order finding, and Grover's search, demonstrate how ONDA simplifies the development of quantum algorithms.

Future research will focus on optimizing the architecture, refining the instruction set, and exploring new programming paradigms enabled by quantum mechanics. The long-term vision is to establish a scalable framework for quantum computing that reduces complexity, improves accessibility, and accelerates practical advancements in the field.

\bibliographystyle{plain}  
\bibliography{references}  

\begin{thebibliography}{10}

\bibitem{Machine_learning}
Jacob Biamonte, Peter Wittek, Nicola Pancotti, Patrick Rebentrost, Nathan Wiebe, and Seth Lloyd.
\newblock Quantum machine learning.
\newblock {\em Nature}, 549(7671):195--202, Sep 2017.

\bibitem{qft_based_addition}
Thomas~G. Draper.
\newblock Addition on a quantum computer, 2000.

\bibitem{Quantum_supremacy}
Frank~Arute et~al.
\newblock Quantum supremacy using a programmable superconducting processor.
\newblock {\em Nature}, 574(7779):505--510, 2019.

\bibitem{QAOA}
Edward Farhi, Jeffrey Goldstone, and Sam Gutmann.
\newblock A quantum approximate optimization algorithm, 2014.

\bibitem{classic_qram}
Vittorio Giovannetti, Seth Lloyd, and Lorenzo Maccone.
\newblock Quantum random access memory.
\newblock {\em Phys. Rev. Lett.}, 100:160501, Apr 2008.

\bibitem{Carry_Save_Arithmetic}
Phil Gossett.
\newblock Quantum carry-save arithmetic, 1998.

\bibitem{demialenzo_quantum_isa}
Francesco Junior~De Gregorio.
\newblock Onda programming language.
\newblock \url{https://github.com/De-Gregorio/ONDA-programming-language}, 2024.
\newblock Accessed: 2024-10-03.

\bibitem{grover}
Lov~K Grover.
\newblock A fast quantum mechanical algorithm for database search.
\newblock In {\em Proceedings of the twenty-eighth annual ACM symposium on Theory of computing}, pages 212--219, 1996.

\bibitem{ibm_evidence_of_utility}
Youngseok Kim, Andrew Eddins, Sajant Anand, Ken~Xuan Wei, Ewout van~den Berg, Sami Rosenblatt, Hasan Nayfeh, Yantao Wu, Michael Zaletel, Kristan Temme, and Abhinav Kandala.
\newblock Evidence for the utility of quantum computing before fault tolerance.
\newblock {\em Nature}, 618(7965):500--505, Jun 2023.

\bibitem{high_dimensional_teleportation}
Wen-Ling Xu, Tie-Jun Wang, and Chuan Wang.
\newblock Efficient teleportation for high-dimensional quantum computing.
\newblock {\em IEEE Access}, 7:115331--115338, 2019.

\end{thebibliography}

\end{document}